\documentclass{article}[15pt]

\usepackage{amsmath,epsfig,color,calc,rotating}
\usepackage{graphics}
\usepackage{wrapfig}
\usepackage{ulem}

\usepackage{natbib}
\bibpunct{(}{)}{;}{a}{}{,}

\begin{document}

\title{
Two-Parameter Characterization of Chromosome-Scale Recombination Rate
\vspace{0.2in}
\author{
Wentian Li and Jan Freudenberg \\
{\small \sl The Robert S. Boas Center for Genomics and Human Genetics}\\
{\small \sl The Feinstein Institute for Medical Research}\\
{\small \sl North Shore LIJ Health System}\\
{\small \sl Manhasset, 350 Community Drive, NY 11030, USA.}\\
}
\date{ }
}
\maketitle  
\markboth{\sl Li and Freudenberg) }{\sl Li and Freudenberg}

\normalsize

\vspace{0.2in}

{\bf ABSTRACT}: 

The genome-wide recombination rate ($RR$) of a species is often described 
by one parameter, the ratio between total genetic map length ($G$) and physical 
map length ($P$), measured in centiMorgans per Megabase (cM/Mb). The value of 
this parameter varies greatly between species, but the cause for these differences 
is not entirely clear. A constraining factor of overall $RR$ in a species, 
which may cause increased $RR$ for smaller chromosomes, is the requirement 
of at least one chiasma per chromosome (or chromosome-arm) per meiosis. 
In the present study, we quantify the relative excess of recombination 
events on smaller chromosomes by a linear regression model, which relates 
the genetic length of chromosomes to their physical length. We find for 
several species that the two-parameter regression, $G= G_0 + k \cdot P$ 
provides a better characterization of the relationship between genetic and 
physical map length than the one-parameter regression that runs through the 
origin. A non-zero intercept ($G_0$) indicates a relative excess of recombination 
on smaller chromosomes in a genome. Given $G_0$, the parameter $k$ predicts 
the increase of genetic map length over the increase of physical map 
length.  The observed values of $G_0$ have a similar magnitude for diverse species,
whereas $k$ varies by two orders of magnitude. The implications of this strategy 
for the genetic maps of human, mouse, rat, chicken, honeybee, worm and yeast 
are discussed.

\newpage

\large

\section*{Introduction}

\indent

The rate of meiotic recombination rate ($RR$), defined as the ratio between 
genetic and physical map length and measured in centiMorgan per Megabase (cM/Mb), 
is known to vary widely between the genomes of different species. As a rule of 
thumb for the human genome, 1cM genetic map length equals 1Mb physical map length, see 
e.g. \citep{collins,ayse}. This rate is about twice as large as the genome-wide 
$RR$ observed in the mouse genome \citep{jensen}, but far less than the $RR$ of 
340cM/Mb that is observed in the yeast genome \citep{ymap1,ymap2}. Understanding of 
these differences in $RR$ between different species is of fundamental importance 
for evolutionary and medical genetics \citep{nachman}. 
In addition to these differences between species, it was also noted
that $RR$ differs between chromosomes within a species, with smaller 
chromosomes showing higher $RR$ \citep{nachman2,broman2,lander, venter,kong,tara}. 
Therefore, species differences in genome-wide $RR$ may be best 
studied under a model that also considers the intragenomic
differences between chromosomes.

From a population genetic perspective, the main role of recombination is 
the production of new combinations of alleles by shuffling of parental haplotypes, 
which increases the efficiency of natural selection in theoretical and 
empirical model systems \citep{maynard,barton,rice,otto}. 
Many recent empirical studies have 
addressed the question at which sites in a genome recombination 
is most likely to occur \citep{petes,mcvean,hey,myers,coop,mancera}.
In this context it was also found that $RR$ evolves extremely 
fast on a kb-scale \citep{ptak,winckler} and that historical recombination 
hotspots are associated with specific gene functions in human, 
which was hypothesized to indicate an influence of natural
selection on hotspot locations \citep{jan,hapmap}. 
When $RR$ is examined at a megabase scale instead of a kilobase scale, 
the evolution of local $RR$ is more constrained \citep{myers}, and differs much less 
between closely related species, such as human and chimpanzees 
\citep{winckler,ptak}.  However, the mechanism behind this conservation of 
$RR$ on the larger scale is unclear. One contributing explanation 
could be the requirement  of a minimal or fixed number of 
chiasmata per chromosome during meiosis to stabilize homologous
chromosome pairs  \citep{mather}.

The question how many chiasmata are exactly required per chromosome 
or per chromosome arm has not been resolved yet and might not have a 
generally valid answer \citep{lynn,laurie}. Nevertheless, these meiotic 
constraints can explain the excess of recombination on shorter chromosomes. 
Consistent with an influence of karyotype on overall recombination rate,
a correlation was found between the number of chromosome arms in a genome 
and the genetic map length \citep{villena}. Altered recombination may 
lead to aneuploidy \citep{hassold,lynn}, which may impose strong selective 
constraints and explain the tight relationship between karyotype structure 
and recombination rate \citep{dumas,villena2}. 

On the other hand, domesticated plants and animals show evidence for 
increased chiasma formation \citep{burtbell}, which suggests that there exist 
additional determinants of genome-wide $RR$ than karyotype. For instance, 
the level of interference in chiasma formation could differ between 
species \citep{broman}. Therefore, it would be useful to apply a formal 
method that separates the contribution of karyotype structure from the 
relationship between physical and genetic map length. This is not 
accomplished by the genome-wide cM/Mb ratio: although the cM/Mb ratio 
is a convenient single parameter measurement, it does not model the 
higher contribution of smaller chromosomes  to the genome-wide $RR$ of a species.

To address this problem and better understand the overall $RR$ of a genome, 
we propose a novel strategy  that explicitly models, 
if and to what extend, the overall $RR$ in a genome 
is influenced by the relative excess of recombination on smaller chromosomes. 
This proposed two-parameter strategy takes into account 
that a certain minimal amount of recombination  is required to 
maintain genome integrity during meiosis \citep{mather} and that a genome therefore 
has minimal genetic map length. This idea becomes more clear, if we 
use a statistical regression framework to compare the proposed strategy with 
the one parameter strategy
that is typically applied to shorter scales than the chromosome-scale. 
Since the one-parameter characterization of $RR$ implies that
genetic length is proportional to the physical length and 
recombination events occur independently on different chromosomes, 
the cM/Mb ratio is the slope of the linear regression of genetic 
lengths of chromosomes over their sequence lengths, with the
requirement that the regression line goes through the origin. 
In our new approach, we drop the requirement that the regression line 
must go through origin by using two parameters to fit the genome-wide 
genetic map information at the chromosomal scale.

From a biological perspective, the  one-parameter model considers
the length of the genetic map of a genome to be  determined by the length of 
the underlying physical map and the species-specific $RR$. Building on this, the  
two-parameter model also includes a separate effect of karyotype structure
that may produce a disportional distribution of recombination events 
over chromosomes of different length. Under the two-parameter model, 
the value of the $y$-intercept quantifies the relative excess of 
recombination events on a hypothetical chromosome with length zero,  
whereas the slope of the regression measures the increase of genetic 
with physical map length in the same way as the one-parameter model. 
Our results show that in human, as well 
as other species, the two-parameter regression provides a much 
better fit for describing the genetic map length of chromosomes.  

\begin{figure}[ht]
  \begin{turn}{-90}
   \epsfig{file=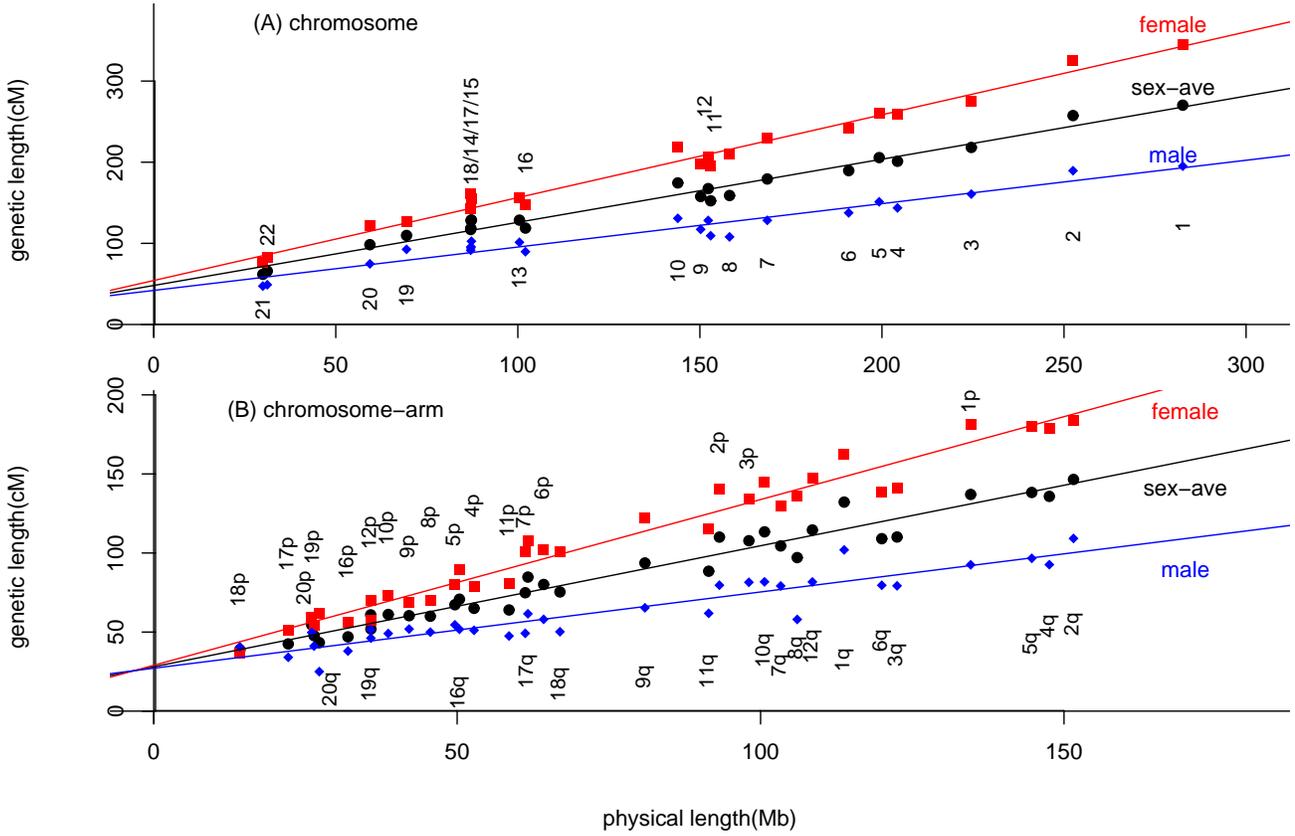, width=11cm}
  \end{turn}
\caption{
\label{fig2}
Two-parameter regression of human genetic length over physical length.
(A) Analysis at the chromosome scale. Female (red), 
male (blue), and sex-averaged (solid circle)
genetic length of each chromosome (in cM) is plotted against its physical
length (in Mb). The least-square regression lines are: $y=54.2 +1.02x$ (female),
$y=42.0+0.52x$ (male), $y=48.1+0.78x$ (sex-average). (B) 
Analysis of metacentric chromosome at the chromosome-arm scale. 
The best fit regression lines are:
$y= 29.0 +1.05x$ (female), $y=27.1+0.48x$ (male),
$y=28.0 + 0.77x$ (sex-average).
}
\end{figure}

\section*{Results}

\subsection*{A two-parameter regression model fits the genetic map length of human chromosomes better than the one-parameter model}

\indent

To look for systematic differences in recombination rate between human 
chromosomes, we started by reproducing the Marey map \citep{chak,mareymap},
a cumulative plot similar to those used in DNA sequence representation
or analysis \citep{wli,grigoriev}, for 22 human autosomes and 34 arms of 
metacentric chromosomes (Appendix Figure A1). 
The chromosome-scale or chromosome-arm scale recombination rate may be  defined
as the slope of a straight line that links the first 
and the last marker. For smaller chromosomes or chromosome arms, the end 
points in the Marey map tend to lie above the line with a slope equal to 1 (cM=Mb), i.e., 
smaller chromosomes(-arms) have larger cM/Mb ratios 
(see also Figure 16 of \citep{lander} and Table 12 of \citep{venter}).

We next regressed the genetic map length of 
chromosomes over their physical map length (Figure \ref{fig2}(A), similar
plot can be found in \citep{housworth}). 
When sex-averaged, female and male genetic lengths 
are fitted separately, the three regression lines are described by:
\begin{eqnarray}
\label{eq1}
G_{ch,sex-ave,human}  &=& 48.1+ 0.78 P \nonumber \\
G_{ch,female,human} &=& 54.2+ 1.02 P \nonumber \\
G_{ch,male,human}   &=& 42.0+ 0.53 P
\end{eqnarray}
The normality assumption of regression residuals was tested graphically
by a QQ-plot (Appendix Figure A2), and the normality condition does 
not seem to be violated.

Equation (\ref{eq1}) shows that the $y$-intercept $G_0$ for female
data is 29\% larger than for male data, whereas the slope $k$ is 92\%
larger. Thus, the different length of the male and female map 
mainly manifests as a different slope and less so as a different $y$-intercept. 
As can be seen from Figure \ref{fig2}(A), all human chromosomes exceed 
the minimal length of 50cM both for the male and the female genetic map. 

To test the robustness of the $y$-intercept value, we added
random noise to the genetic map length and repeated
the regression analysis. The histogram of 50000 $y$-intercepts from 
this procedure is shown in 
Appendix Figure A3.  Although  values of $G_0$ range from 
35 to 60, they are all far from zero.

We next repeated the analysis using chromosome arms instead of
full chromosome as separate data points (Figure \ref{fig2}(B)).
This leads to the regression equations:
\begin{eqnarray}
\label{eq2}
G_{arm,sex-ave,human}  &=& 28.0+ 0.77 P \nonumber \\
G_{arm,female,human} &=& 29.0+ 1.05 P \nonumber \\
G_{arm,male,human}   &=& 27.1+ 0.48 P.
\end{eqnarray}

The $y$ intercepts at chromosome arm-scale regression is now reduced
to somewhat more than half of the intercept at the full chromosome scale.
This reduction shows that cytogenetic constraints exert a smaller 
influence on the chromosome-arm scale than on the full chromosome scale.

Several methods can be used to show that the two-parameter regression model 
fits the data better than the one-parameter regressions. 
To this end, we first compared the coefficient of determination $R^2$ which is the 
proportion of variability explained by the regression model. The observed $R^2$
values of the one-parameter regression 
range between 0.48 and 0.87, whereas the $R^2$
values of the two-parameter regressions range between 0.86 and 0.98 (Table 1),
indicating that the two-parameter regression explains more of the variability 
in the data. 

We further cast the comparison between the one- and two-parameter 
regression as a model selection problem. 
Two such model comparison strategies are 
provided  by the Akaike information criterion (AIC) \citep{aic}
and Bayesian information criterion (BIC) \citep{bic}. 
Both AIC and BIC values for the two-parameter regression model
are smaller than those for the one-parameter model, indicating 
a better statistical model (Appendix Table A1).

Finally, we tested the null hypothesis that $G_0$ is zero. 
The $p$-values in this test range between $10^{-13}$ and $10^{-8}$ (Table 1). 
Because the null hypothesis is that all chromosomes have the same RR, 
a simulated distribution of $G_0$ can be obtained from the regression 
over data that are obtained by the permuting of chromosome-specific
cM/Mb ratios, while leaving the physical chromosome length unchanged. 
Out of 50000 such permutations, only two showed a
$G_0$ value that is larger than the observed value of 48.1 
(for sex-averaged full chromosome data), corresponding to
a $p$-value of $4 \times 10^{-5}$.
To summarize, all evaluation methods support the conclusion that the
two-parameter regression model is better than the one-parameter model.

\begin{table}[ht]
\begin{center}
\begin{tabular}{r|c|c|c}
\hline
 % & \multicolumn{2}{c|}{two-parameter regression} & 
 %\multicolumn{2}{c|}{one-parameter regression} & $p$-value for \\
 & two-parameter & one-parameter &  $p$-value for  \\
& $R^2$ &  $R^2$ &   testing $G_0=0$ \\
\hline
human chromosome, sex-averaged  
 & 0.976 &  0.817 & 3.1 $\times 10^{-10}$   \\
 female 
 & 0.984 &  0.866 & 8.6 $\times 10^{-11}$  \\
 male 
 & 0.942 &  0.694 & 1.2 $\times 10^{-8} $  \\
\hline
human chromosome arm, sex-averaged  & 0.955 & 0.775 &  9.3 $\times 10^{-13}$   \\
 female  & 0.964 &  0.861      &  7.1 $\times 10^{-11}$   \\
 male  & 0.864 &  0.480        &  7.7 $\times 10^{-11}$   \\
\hline
\end{tabular}
\end{center}
\caption{\label{table1}
Comparison of the two-parameter and one-parameter regression models for human genetic
length, at the chromosome scale (22 data points) and the chromosome-arm scale
(34 data points):
coefficient of determination ($R^2$) for 2- and 1-parameter regressions, and
$p$-value for testing the null hypothesis of zero $y$-intercept.  }
\end{table}

As the deCode data were published more than six years ago, we 
further tested the chromosome-scale regression strategy on a more recent dataset,
the Rutgers Map v.2 \citep{tara}. The regression lines are
$G= 53.33 + 0.87 P $ (sex-average), $ G= 50.29 + 1.19 P $ (female), 
and $ G= 57.58 + 0.57 P$ (male), respectively.
These results are consistent with the parameter estimations in
Eq.(\ref{eq1}), again showing that male and female data differ more
in the slope than in the $y$-intercept.

Two other quantities can be derived from $G_0$ that help to interpret
the $y$-intercept parameter. The first is the physical length $P_{min}$ on the regression line
that corresponds to a specified minimum genetic length $G_{min}$ such
that $G_{min}= G_0+ k P_{min}$. If we set $G_{min}=$ 50cM, then we obtain 
$P_{min}=$ 2.45Mb  for sex-averaged chromosome data.
One may assume that for any hypothetical chromosome with $P < P_{min}$, 
its genetic length $G$ remains constant at 50cM and does not
decrease for shorter chromosome length.  As the second quantity of interest, 
we define the percentage of genetic length that is explained
by the inclusion of $G_0$ into the model as: 
$\alpha= 22 G_0/(22 G_0 + k \sum_{i=1}^{22} P_i)$.
For the sex-averaged data, we find that $\alpha = 31\%$ of variability
is explained by the $y$-intercept. This value can also be
obtained from the decomposition 
of $RR= \sum_{i=1}^{22}G_i/\sum_{i=1}^{22}P_i = 22 \cdot G_0/\sum_{i=1}^{22} P_i + k$:
1.13= 0.35 + 0.78, because 0.35/1.13 = 31\%.  The relatively large
percentage value once again highlights the importance of the $y$-intercept $G_0$
for modeling chromosome-scale recombination rate in human. 

\begin{figure}[th]
  \begin{center}
  \begin{turn}{-90}
   \epsfig{file=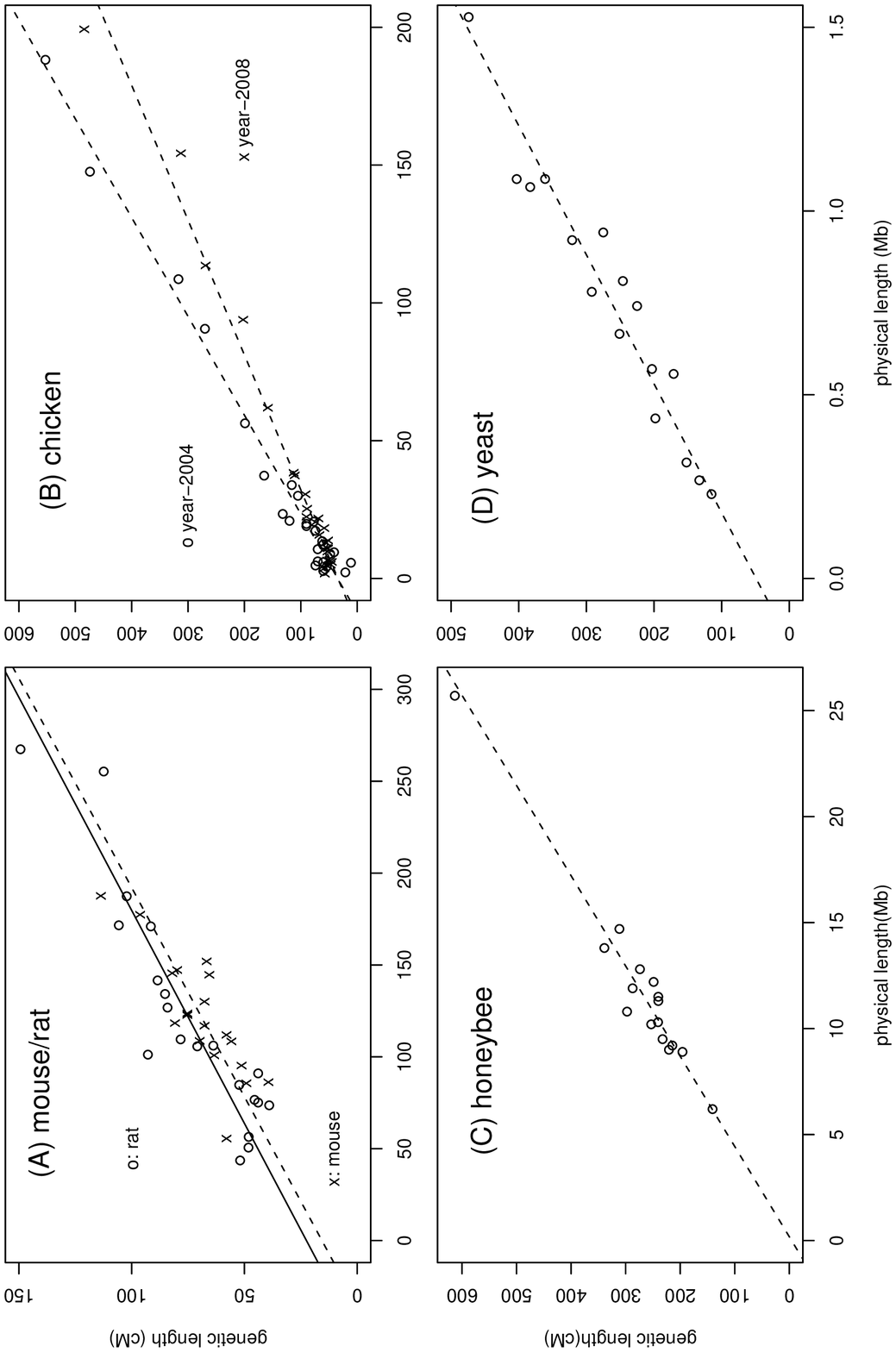, width=10cm}
  \end{turn}
  \end{center}
\caption{
\label{fig3}
The genetic length (in cM) vs. physical length (in Mb) plotted for
five genomes: (A) mouse (crosses) ({\sl Mus musculus}) 
and rat (circles) ({\sl Rattus norvegicus}). Source: Table 1 of \citep{jensen}. 
The regression lines are: $y=15.62 +0.44x$ (mouse),
$y=22.49 +0.43x$ (rat).
(B) chicken ({\sl Gallus gallus}). Source: old data (year 2004, circle) is
from  supplementary Table A2 of \citep{chicken}; new data (year 2008, cross)
is from Table 1 of \citep{chicken2}.
The regression line is: $y=34.68 + 2.79x$ (old data) and
$y= 34.23 + 2.04 x$ (new data).
(C) honeybee ({\sl Apis mellifera}). Source: Table 2 of \citep{beye}.
The regression line is $y= -4.22 +23.49x$.
(D) budding yeast ({\sl Saccharomyces cerevisiae}). Source: 
{\sl http://downloads.yeastgenome.org/chromosomal\_feature/SGD\_features.tab}.
The regression line is $y=49.12 +287.74x$.
}
\end{figure}

\subsection*{Different intercept but similar slope in the two-parameter 
regression models for rat and mouse chromosomes}

\indent

Both rat ({\sl Rattus norvegicus}) and mouse ({\sl Mus musculus})
genome are known to have lower recombination rates 
than human \citep{jensen}, with rat having a higher overall $RR$ than mouse. 
The rat genome has a roughly equal physical map length, but 
contains one more chromosome (n=20) than the mouse genome
(n=19). Furthermore, rat chromosomes show a greater heterogeneity
in their physical length and 
one may hypothesize that these karyotype differences contribute 
to the somewhat higher $RR$ in rat (0.62 cM/Mb vs. 0.57 cM/Mb in mouse). 
The regression models of the sex-averaged genetic length of rat and mouse 
chromosomes over their sequence lengths (Figure \ref{fig3}(A)) are: 
\begin{eqnarray}
\label{eq3}
G_{ch,sex-ave,rat}  &=& 22.49+ 0.43 P \nonumber \\
G_{ch,sex-ave,mouse}  &=& 15.62+ 0.44 P.
\end{eqnarray}
These models display a similar slope and the different overall $RR$
of rat and mouse mainly manifests as a different intercept value $G_0$.

Testing $G_0=0$ for the rat genome is significant
($p$-value= 0.0012), whereas testing $G_0=0$ for mouse genome (the fitted
$G_0$ value for mouse is 69\% of that for rat) is not
significant ($p$-value = 0.11).
AIC/BIC calculation confirms that the two-parameter regression is a
convincingly better model for rat than the one-parameter regression,
whereas this barely holds for the mouse data (Appendix Table A1).
Thus, the mouse genome displays a 
non-significant excess of recombination on smaller chromosomes, 
which is consistent with the smaller variation of chromosome size 
in the mouse genome.  The greater $y$-intercept for the rat genome
supports the hypothesis that cytogenetic factors contribute more
to the genetic map length of rat than mouse.

Because the rat karyotype consists of both meta- and acrocentric chromosomes, 
we repeated the analysis after splitting all metacentric rat chromosome 
into two parts, based on the location of the centromere. Different 
from the human genome, for the rat genome this mainly affects the 
smaller chromosomes, which are often metacentric. The regression line 
is now  described by $G = 7.48 + 0.52P$ and testing the intercept is 
still significant ($p$-value=0.024), though at a less stringent level. 
Thus, at the scale of chromosome-arms, the likelihood of crossovers 
in an interval is more determined by its physical length and less by influenced 
by any obligate recombination requirements.

\subsection*{Recombination rate of small
and large chromosomes in the chicken genome}

\indent

The chicken ({\sl Gallus gallus}) genome consists of both large 
(macro-) and small (micro-) chromosomes \citep{chicken,smith}, with
length ranging from a few Mb to close to 200Mb.  
The two-parameter regression model for the chicken genetic 
data in Figure \ref{fig3}(B) leads to:
\begin{equation}
\label{eq_chicken1}
	G_{ch,sex-ave,chicken}= 34.68 + 2.79 P.
\end{equation}
In this regression model, the non-zero intercept is significant with a a $p$-value 
of $1.33 \times 10^{-7}$ and there is a considerable difference of AIC/BIC for 
the one- and two-parameter regression
favoring the two-parameter model (Appendix Table A1). 
Both coefficients of determination for the 2- and 1-parameter regressions attain 
a high value: 0.98 and 0.93 respectively.  A reason that the 1-parameter 
regression only marginally reduces the $R^2$ value is given by the 
fact that larger chromosomes contribute much more to the total variance, 
which is equally well captured by the 1-parameter model. Thus, two of the three 
methods confirm a relative  excess of recombination on short chromosomes.

However, the orders of magnitude difference between the size of chicken 
chromosomes raises the question of the robustness of the regression.
From \citep{chicken} and Appendix Figure A4, it is clear that 
the genetic length reaches a plateau at the level of 50cM for microchromosomes
smaller than 8Mb.
When chromosomes below a certain length threshold are 
discarded from the regression analysis, the $y$-intercept value
changes slightly, but not dramatically. For example, if the length
thresholds for removal are 8Mb and 25Mb, $G_0$ for Eq.(\ref{eq_chicken1})
decreases to 32.95 and 31.88.
When the regression model is only fitted to the five 
largest chromosomes (longer than 50Mb), the model parameters are $G = 26.22 + 2.84 P$. 
On the other hand, if we remove the largest five chromosomes,
the regression line is $G = 31.86 + 3.01 P$.

To see how the quality of the map distance measurements may influence
these results, we next looked at the recently updated chicken map \citep{chicken2},
which contains more genetic markers and higher marker density. 
Applying the two-parameter regression leads to
\begin{equation} 
\label{eq_chicken2}
	G_{ch,sex-ave,chicken}= 34.23 + 2.04 P.
\end{equation} 
As can be seen, the overall reduction of $RR$ as compared to
the older map \citep{chicken} mainly manifests as reduced
estimate of $k$, whereas the estimate of $G_0$ remains almost unchanged.

\subsection*{Exceptionally high recombination on the largest honey bee 
chromosome leads to a better fit of the one-parameter than the two-parameter model}

Notably, the two-parameter regression does not provide a better fit for 
the genetic map data from honey-bee({\sl Apis mellifera}) \citep{beye}
than the one-parameter model. When plotting 
the genetic length over physical length (Figure \ref{fig3}(C)),
the $y$-intercept of the regression line does not significantly differ from zero
($p$-value $=0.81$):
\begin{equation}
G_{ch,sex-ave,bee}= -4.22 + 23.49 P.
\end{equation}
The coefficient of determination for both the two- and one-parameter regression
is around 0.95. In contrast to other genomes, AIC/BIC analysis
favors the one-parameter regression model (Appendix Table A1).

As can be seen from Figure \ref{fig3}(C), the longest chromosome
(chromosome 1) is four times the length of the shortest chromosome, and
the regression result may depend on the presence of this ``outlier". 
To check this possibility, we repeated the analysis after chromosome 1 
was removed, which led to the regression equation: $G= 28.71 + 20.36 P$.
However, also in this model,  testing $G_0=0$ is not significant ($p$-value=0.37),
both the two- and one-parameter regressions exhibit similar coefficient
of  determination ($R^2=$ 0.80, 0.79), and the zero-intercept regression
is still the better model according to AIC/BIC analysis (Appendix Table A1).
Therefore, different from other species, the honey-bee genome does not 
display any significant excess of recombination on smaller chromosomes.

\subsection*{Two-parameter regression at much shorter length scales:
the example of budding yeast}

Yeast ({\sl S. cerevisiae}) has been extensively used to study the molecular 
machinery of recombination and it has a much smaller ($\sim$ 12Mb)
and more compact genome \citep{yeast}. 
Although the physical length of yeast chromosomes only ranges from 
200kb to 1.5Mb, their genetic length is between 100 and 500cM, 
even longer than the genetic length of human chromosomes. 
The best fitting regression line for the yeast genetic map is 
(Figure \ref{fig3}(D)):
\begin{equation}
G_{ch,sex-ave,yeast} = 49.12 + 284.74 P.
\end{equation}
The non-zero $y$-intercept is significant ($p$-value = 0.009).
The two-parameter regression is superior to the one-parameter model
as judged by AIC/BIC (Appendix Table A1).
The value of $y$-intercept, 49.12 cM  is very close to 50cM which 
corresponds to almost one crossing over event for a hypothetical chromosome 
of physical length of zero.

The extremely high recombination rate in the yeast genome is surprising.
From the molecular perspective,
one can speculate about various hypotheses, such as a different meiotic
regulatory system which makes a denser spatial distribution of
chiasmata possible, a lack of secondary chromatin structure as
compared to higher organisms so that the actual physical distance
between two locations on chromosome is more or less equal to the
linear sequence distance, or the lack of other supporting mechanism
to hold chromatids together so that more chiasmata per chromosome arm
are required for proper chromosome segregation. On the other hand, 
the $y$-intercept of the regression has a similar magnitude as 
that observed for higher organisms, indicating a similar relative excess 
of recombination on smaller chromosomes.

\subsection*{The difference between the central gene cluster
and telomeric regions in worm genome is due to a difference in $G_0$}

\begin{figure}[th]
  \begin{center}
  \begin{turn}{-90}
   \epsfig{file=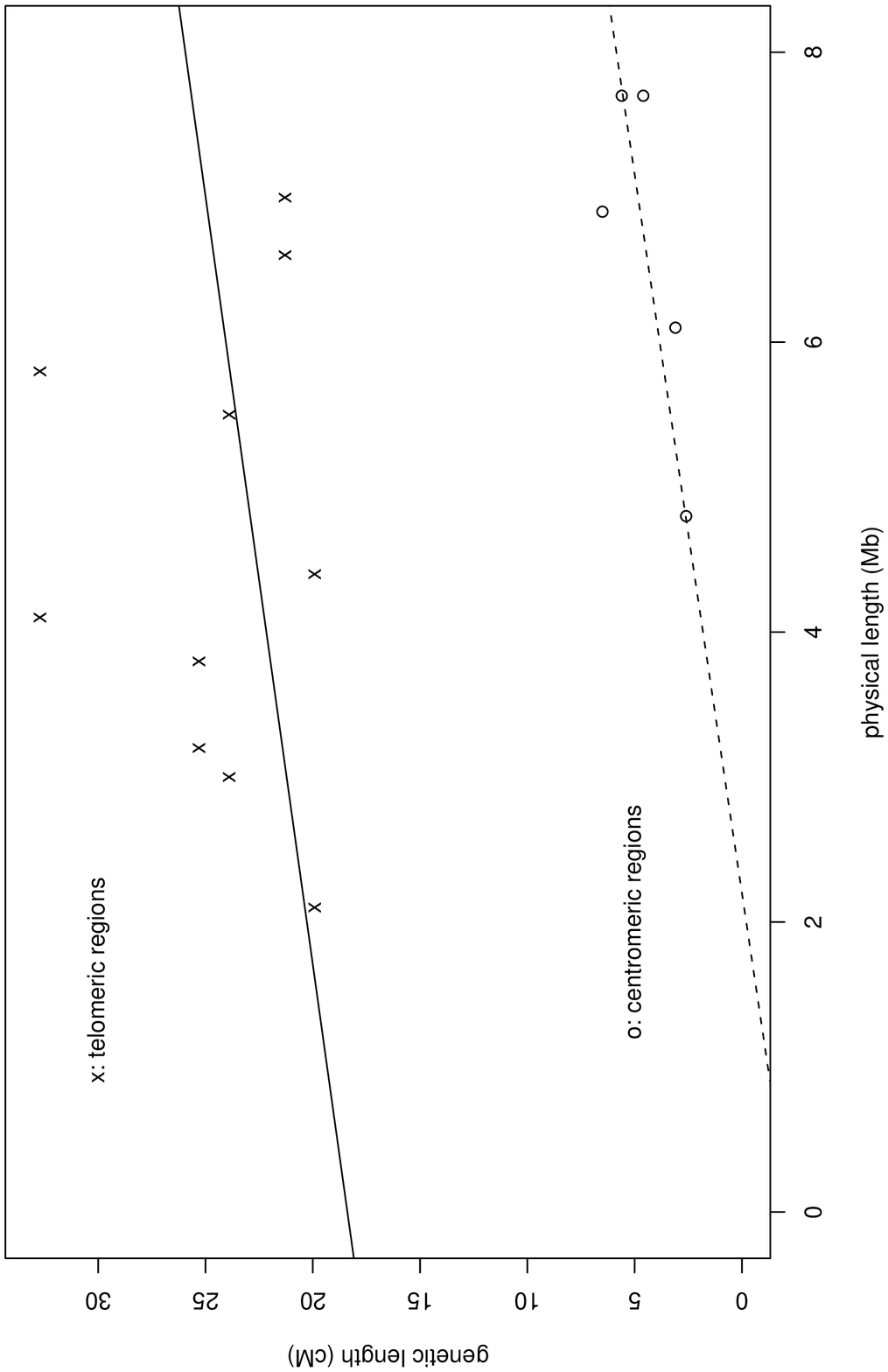, width=8cm}
  \end{turn}
  \end{center}
\caption{
\label{fig4}
The cM-Mb plot using the physical and genetic length of central 
gene clusters (5 data points) and distal arms (10 data points) of
five worm ({\sl Caenorhabditis elegans}) chromosomes 
(Table 1 of \citep{barnes}).  The best fitting regression lines 
are $y=18.39 +0.94x$ for the distal/telomeric arms (crosses),
and $y=-2.22+1.01x$ for the central gene cluster regions (circles).
}
\end{figure}

Finally,  we used genetic map data from the worm {\sl C. elegans} to 
show that the two-parameter regression strategy can also be useful 
to compare different regions within a genome.  The chromosomes 
of {\sl C. elegans} are unusual, because discrete centromeres are 
missing and the chromosomes are holocentric, i.e. microtubules 
attach at many sites for chromatid segregation \citep{tyler-smith}. 
Accordingly, the Marey map analysis of the worm genome indicates 
that each worm chromosome can be partitioned in three regions: the 
central gene-rich region with a low recombination rate and two 
distal telomeric regions with high recombination rates \citep{barnes}.
Therefore, we separately performed the regression analysis of 
genetic length over physical length for these two types of regions 
(Figure \ref{fig4}).
The fitted regression coefficients are:
\begin{eqnarray}
\label{eq6}
G_{central,worm} &=& -2.22 + 1.01 P \nonumber \\
G_{distal,worm} &=& 18.39 + 0.94 P
\end{eqnarray}

Within the single parameter framework without the intercept term
$G_0$, the two
types of regions would have a very different cM/Mb ratio:
4.57 for telomeric regions, 0.68 for central regions.
However, when allowing non-zero $G_0$ value, the two regions 
display similar slope values, 1.01 and 0.94. This indicates
a constant excess of recombination in the distal region
as compared to the central region in {\sl C. elegans},
which is combined with a similar incremental cM/Mb ratio.
Thus, after accounting for a fixed amount of recombination 
in a distal chromosome region, the likelihood of any additional 
recombination depends in similar strength on physical length 
in distal regions and central regions.

As a note of caution, one may point out that the regression coefficient 
in Eq.(\ref{eq6}) is obtained from only a few data points. 
Nevertheless, further regression diagnostics 
supports our conclusion. For example, testing $G_0=0$ is
significant for the distal regions ($p$-value= 0.006),
but not significant for central regions ($p$-value=0.57).
AIC/BIC analyses lead to the same conclusion (Appendix Table A1).

\section*{Discussion and Conclusions}

\indent

Our results show that instead of the simpler genetic-to-physical length 
ratio, the relationship between the physical and genetic map length
at chromosome scale is better described by a statistical model that contains 
a second parameter $G_0$, which is the $y$-intercept of the regression 
of genetic map length over the physical chromosome length.  A conceptually similar
approach was used earlier in measuring the genome-wide recombination rate 
of a species by counting the chiasmata on each chromosome in excess of one \citep{burtbell}.

The consideration of this intercept parameter is important, because 
karyotype structure has been established as an important
determinant of genome-wide $RR$ \citep{villena,coop} and
smaller chromosomes display higher $RR$ \citep{lander,chicken}.
Our proposed two-parameter model provides a formal expression 
of this size dependency of $RR$: $RR= G/P= k + G_0/P$,
i.e., a constant term $k$ plus a second term
that increases for smaller chromosome sizes $P$ (if $G_0$ is positive). 
This is what we observe for human, mouse, rat, chicken and yeast genomes. 
When writing $G_0$ as $G_0= G- kP$, the $y$-intercept measures the amount of recombination
after the physical map length has been accounted for. Therefore, one
would expect that the total map length $G$ of a chromosome increases
by $G_0$ after splitting it up into two separate parts.
In fact, this has already been quantitatively observed for 
the experimental alteration of yeast chromosome I \citep{kaback}.

When comparing $RR$ between species, the usage of $k$ instead of the genome-wide cM/Mb ratio 
will reduce the influence of karyotype differences on the result. 
This was also the intention behind the counting of chiasmata per chromosome
in excess of one \citep{burtbell}. In our study, the order of species remains
unchanged, whether ranked by $k$ or by cM/Mb ratio. 
However, due to the different values of $k$, we cannot use a single regression
line to model the genetic-physical length relationship across species.
Thus, a molecular mechanism must exist
that drives, within a particular species, the proportional increase
of genetic over physical map length. This mechanism might typically act 
with weaker strength in larger genomes, which could contribute to the
inverse correlation between genome size and $RR$ \citep{lynch}.

Among mammals it was furthermore found that $RR$ is more similar
for more closely related species \citep{dumont}, which could
be partly due to their similar karyotype. It might be interesting to test
where in the phylogenetic tree the signal might be altered,
when using $k$ instead of the genome-wide cM/Mb ratio. 
In this context, it is also important that genome-wide $RR$
typically differs between genders and individuals
\citep{broman2,kong,kong2,cheung,petkov}. The biological
factors that were invoked as possible explanations, such as
differences in synaptonemal complex formation or crossover
interference, may be more plastic than karyotype structure.
The respective strength of these factors could also contribute
to species differences and may be better measured by
using $k$ than by using the genome-wide cM/Mb ratio.

If $k$ would equal to zero with the obligate chiasma requirement holding true,
then $G_0$ were required to be 50cM. 
This pattern can be observed for female Opossum ({\sl Monodelphis domestica}), where
each chromosome acquires exactly one crossover near one of its
telomeres (see \citep{opossum1} and Appendix Figure A5). Similarly, very small chromosomes
may always acquire exactly one crossover, despite reduced 
chromosome size, as seen for the  microchromosomes
in the chicken genome. In order to predict the transition
from this plateau to the linear regression, we derived the minimum physical
length parameter $P_{min}$ from a given $G_{min}$ and the estimated
regression parameters.  Note that if both
physical and genetic length are measured as those
in excess of $P_{min}$ and $G_{min}$, their ratio is exactly
equal to $k$: 
$(G-G_{min})/(P-P_{min})= (G-G_{min})/(P - (G_{min}-G_0)/k)=
(G-G_{min})/( (kP+G_0 - G_{min})/k)=k$.

Because reduced recombination may result in aneuploidy of smaller
chromosomes \citep{warren,brown}, it is conceivable that the length
of smaller chromosomes could influence genome-wide $RR$ by introducing 
a lower bound for the propensity for chiasma formation in a species.
Our analysis supports the size of the smaller chromosomes as
a strong determinant of genome-wide $RR$ for the six genomes studied
in this paper (Appendix Figure A6). In log-log
scale, the correlation coefficient between $RR$ and the
shortest chromosome length is $-0.92$ ($p$-value= 0.008).
If the recombination rate is measured by $k$, in log-log scale
the correlation coefficient is $-0.91$ ($p$-value=0.01). 
This correlation is nearly as strong as the reported correlation between
$RR$ and the total physical length for over 100 genomes
(cc= $-0.99$, $p$-value= 0.0003 on log-log scale) as reported 
in \citep{lynch}. Obviously, data on more species are needed for 
a more conclusive analysis. Nevertheless, it may be interesting to point out
that the genome with the lowest known recombination rate, Opossum,
lacks any short chromosome \citep{opossum1,opossum2}.

Obviously, any genome-wide analysis relies on the availability
of high quality data. We are convinced that the data used in this
study are of sufficient quality to study recombination on the
chromosomal scale. However, some error might be introduced by
the fact that the used genetic maps are not perfect and, in
particular for telomeres, missing some data. This can be seen
for the chicken genomes, where the two chromosomes fall below the 
minimum genetic length of 50cM in the older map and climb to about
50cM in the newer map (Appendix Figure A4).  
Data selectively missing crossovers at the telomeres might lead
to an underestimation of $G_0$ in the regression model.

We note that we restricted our analysis to chromosomes or chromosome-arms.
If the genetic length is regressed over the length of much smaller
regions, the coefficient of determination $R^2$ is expected to
be much lower due to a mixture of recombination hot- and cold- spots. 
From a biological perspective, we also would not expect a positive 
$G_0$ value in such a regression, because there is no requirement
for a Mb-sized region to have at least one chiasma to maintain meiotic
integrity.

In summary, we find that the introduction of the $G_0$ 
parameter helps to understand the recombination rate differences 
between species, because it separates the effect of the
requirement for at least 
one chiasma formation on smaller chromosomes from the factors that determine the 
amount of recombination on larger chromosomes. More specifically, 
the partitioning of chromosome-scale recombination rate leads
to the following list of conclusions: 
i) human male-female $RR$ differences disproportionally affect larger 
chromosomes;
ii) the higher recombination rate in the rat 
genome as compared to the mouse genome is likely to be caused by 
the higher number of smaller chromosomes  that constitute the rat karyotype; 
iii) both  chicken micro- and macro-chromosomes display a high 
$RR$ and the extraordinarily high $RR$ of some micro-chromosomes 
does not lead to an extraordinary excess of recombination on smaller chromosomes; 
iv) the honey-bee genome does not display any significant
excess of recombination on smaller chromosomes; v) yeast displays 
a relative excess of recombination on smaller chromosomes that 
is similar to higher organisms, despite its outstandingly high overall 
recombination rate;
vi) recombination of the worm genome mainly occurs in telomeric regions 
and given one recombination per chromosome arm, the likelihood of a 
second recombination is determined by physical map length.
These examples demonstrate that the proposed statistical framework 
allows to pinpoint differences of genomic recombination rate, 
which should be useful for the further study of genome-wide 
recombination rate as a quantitative trait of fundamental importance.

\section*{Materials and methods}

\subsection*{Genetic map data}

The human genetic map was obtained from \citep{kong} that uses 
5136 microsatellite markers with 1257 meiotic events,
and is estimated from pedigree data
(Supplementary Table E: \\
{\sl http://www.nature.com/ng/journal/v31/n3/suppinfo/ng917\_S1.html}).
The rat ({\sl Rattus norvegicus}) and mouse ({\sl Mus musculus})
genetic map data were obtained from Table 1 of \citep{jensen}, based on 
2305 markers in rat and 4880 markers in mouse.  The two chicken 
({\sl Gallus gallus}) genetic maps were obtained from Supplementary 
Table S2 of \citep{chicken}, which is built from 1471 markers, 
and from Table 1 of \citep{chicken2} built from 9258 markers.  
The honey bee ({\sl Apis mellifera}) genetic map was obtained
from Table 2 of \citep{beye} based on 1500 markers.  The budding yeast 
({\sl Saccharomyces cerevisiae}) genetic map was downloaded from \\
{\sl http://downloads.yeastgenome.org/chromosomal\_feature/SGD\_features.tab}.
The worm ({\sl Caenorhabditis elegans} physical and genetic 
lengths of central ``gene clusters" and distal ``arms"
were obtained from Table 1 of \citep{barnes} based on
168 markers.

\subsection*{Measuring how good a linear regression is by 
coefficient of determination}
Regression analyses were carried out by the $lm()$ subroutine in
$R$ statistical package. For genetic lengths, $\{ G_i \}$
($i=1,2, \cdots n$, e.g., $n=22$ for the chromosome-scale
regression and $n=34$ for the chromosome-arm-scale regression),
one can regress them over sequence lengths $\{ P_i \}$
($i=1,2, \cdots n$) allowing $y$-intercept (non-zero $G$
when $P$ approaches 0): 
\begin{equation}
\label{2p}
G= G_0 + k P, 
\end{equation}
or, without the $y$-intercept ($G$ approach 0 as $P$ approaches 0):
\begin{equation}
\label{1p}
G= kP.
\end{equation}

How good a linear regression model fits the data can be measured
by the coefficient of determination $R^2$, which is the proportion 
of variability that is explained by the model. More specifically, if
$SS_{tot} = \sum_{i=1}^n (G_i - E[G_i])^2 $ is the total
sum of squares of the genetic lengths of chromosomes,
the term $SS_{err}= \sum_{i=1}^n (G_i - G_0 - k P_i)^2$
for allowing non-zero $y$-intercept,
or the term,
$SS_{err}= \sum_{i=1}^n (G_i - k P_i)^2$ for
not allowing $y$-intercept,
is the residual sum of squares (RSS), then
\begin{equation}
R^2 = 1 - \frac{SS_{err}}{SS_{tot}}
= 1 - \frac{RSS}{SS_{tot}}
\end{equation}

\subsection*{Model selection by Akaike information criterion}

Akaike information criterion (AIC)\citep{aic} of a statistical model
is defined as $2p-2log(L)$ where $p$ is the number of parameters
in the model, and $L$ is the maximum likelihood estimated from
the data. Similarly, Bayesian information criterion
(BIC)\citep{bic} is defined as  $\log(n)p-2log(L)$, where
$n$ is the number of samples used to calculate the likelihood. 
For linear regression, AIC/BIC is related to the
residual sum of squares (RSS) according to \citep{ripley} by:
\begin{eqnarray}
AIC &=& 2p + n \log(RSS/n) \nonumber \\
BIC &=& \log(n) p + n \log(RSS/n)
\end{eqnarray}
where $n$ is the number of sample points for the regression analysis.
Between two statistical models that are fitted to the same dataset, the
model with a smaller AIC/BIC value is considered to be better 
than the model with a larger AIC/BIC value.

For the comparison between the two- and one-parameter regressions,
we have:
\begin{eqnarray}
AIC_2-AIC_1 &=&  2 - n \log \frac{RSS_1}{RSS_2} \nonumber \\
BIC_2-BIC_1 &=&  \log(n) - n \log \frac{RSS_1}{RSS_2} 
\end{eqnarray}
If the second term, $n \log( RSS_1/RSS_2)$, is larger than two (for AIC,
or $\log(n)$ for BIC),
then the two-parameter regression can be seen as the better model than
the single-parameter regression.

\subsection*{Quantities derived from $G_0$ }

The linear relationship between $G$ and $P$ cannot extend to the
physical length of zero, if the $y$-intercept is greater than zero
and the obligate chiasma requirement holds. 
Therefore, a point $P_{min}$ must exist below which genetic
map length remains constant at $G_{min}$, independent
from the actual physical map length of a chromosome.
We can define this transition point as follows:
$P_{min}$ is the physical length for which the regression line 
crosses the horizontal line defined by the  minimum genetic length 
$G_{min}$, thus $P_{min}= (G_{min}-G_0)/k$.

Another derived quantity is the genome-wide percentage of 
genetic length that is  explained by
$G_0$. For a single chromosome ($i$), this percentage is $\alpha_i \equiv  G_0/(G_0+k P_i)$.
For the whole genome, it is $\alpha \equiv n G_0/( n G_0 + k \sum_i P_i)$,
where $n$ is the number of chromosomes. This definition of $\alpha$ is
valid only when the $y$-intercept is positive ($G_0 > 0$ ).

\vspace{0.5in}

\section*{Acknowledgements}

We thank Alejandro Morales for participating in the initial stage of
this work, Tara Matise, Zhiliang Hu, Hong Ma, Oliver Clay for discussions
and the anonymous reviewers for their valuable comments and suggestions.
J.F. was supported by a NARSAD Young Investigator award.

\newpage

\section*{Appendix}

\setcounter{figure}{0}
\setcounter{table}{0}

\renewcommand{\thefigure}{A\arabic{figure}}
\renewcommand{\thetable}{A\arabic{table}}

\section{Marey Map for human chromosomes (Fig.A1) }

\begin{figure}[ht]
  \begin{center}
  \begin{turn}{-90}
   \epsfig{file=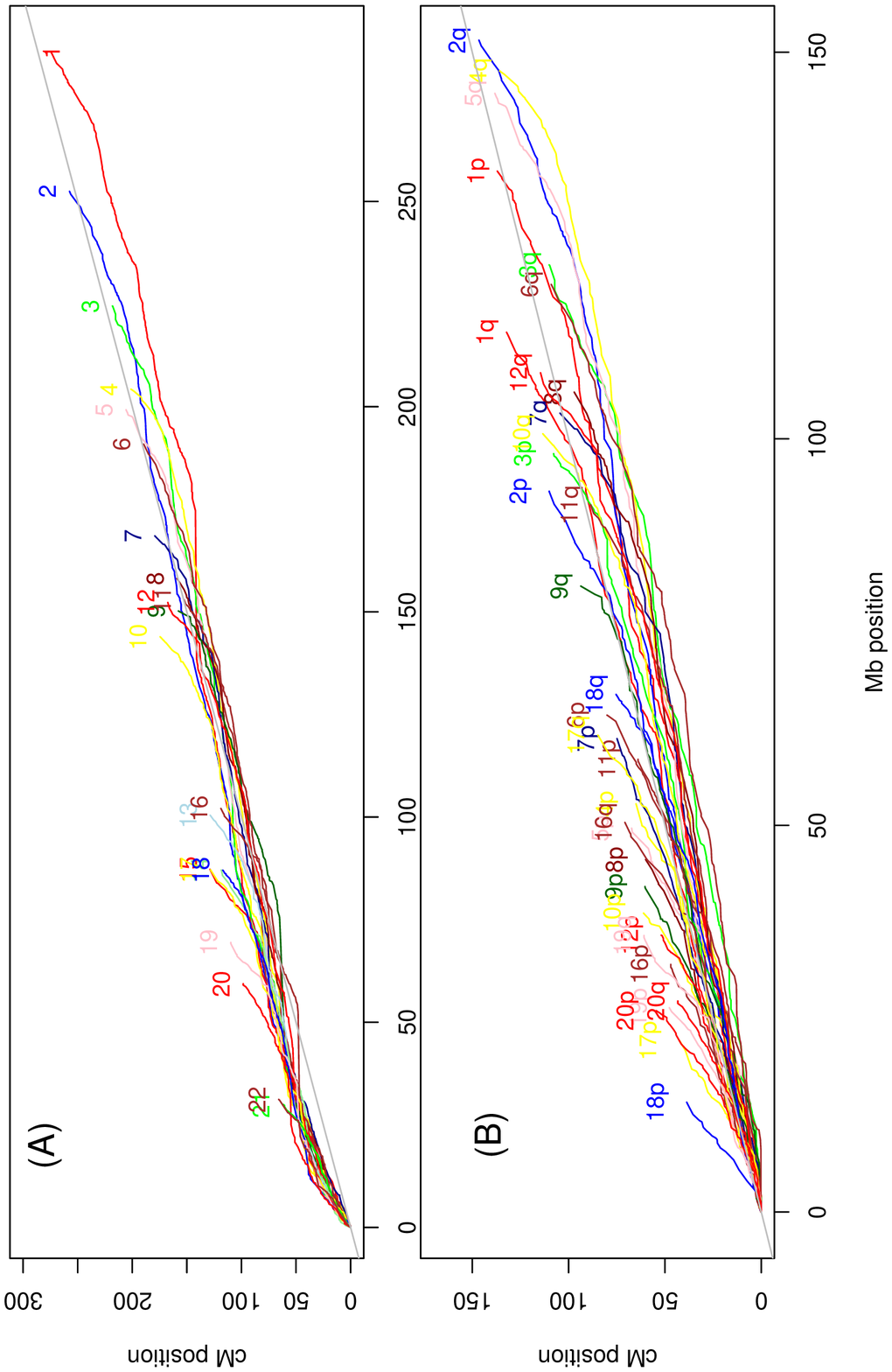, width=10cm}
  \end{turn}
  \end{center}
\caption{
Marey map for human genetic data (Supplementary Table E of (Kong et al., 2002),
with $y$-axis showing the genetic distance (in cM) from the first marker 
to the current marker, and $x$-axis showing the physical distance (in Mb).
(A) Each line traces a chromosome (chromosome name is shown as label).
(B) Each line traces an arm of a meta-centric chromosome 
(p- and q-arms are shown as label).
The straight line indicates cM=Mb.
}
\end{figure}

\newpage

\section{Checking the normality assumption of the regression (Fig.A2) }

\begin{figure}[ht]
  \begin{center}
  \begin{turn}{-90}
   \epsfig{file=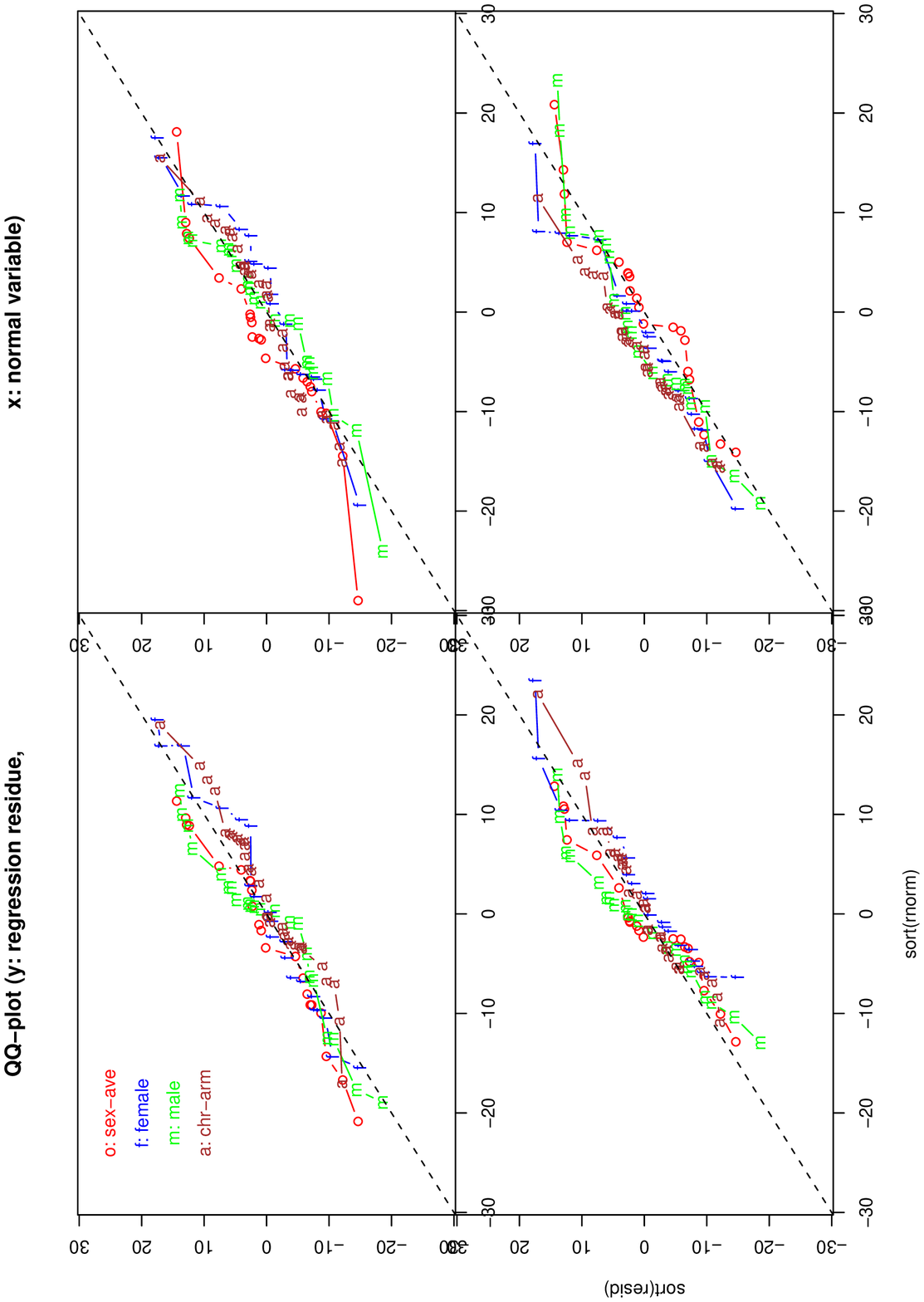, width=10cm}
  \end{turn}
  \end{center}
\caption{
QQ-plots of the residuals of regression ($\epsilon= G- G_0- kP$) against
simulated normal variables. The variance of the normal variables is chosen
to be equal to that of the regression residuals.  Due to the small 
number of sample points (22 for human chromosomes, 34 for human chromosome arms), 
four sets of normal random variables were generated to test the robustness. 
In each QQ-plot, ``o" denotes the regression for the sex-averaged map of human 
chromosomes, ``f" for the human female map data, ``m" for human male map
data, and ``a" for sex-averaged human map data for chromosome arms 
(see Eqs.(1,2)). 
}
\end{figure}

\newpage

\section{Testing the robustness of regression result by
adding random  noise to the genetic map (Fig.A3) }

\begin{figure}[ht]
  \begin{center}
  \begin{turn}{-90}
   \epsfig{file=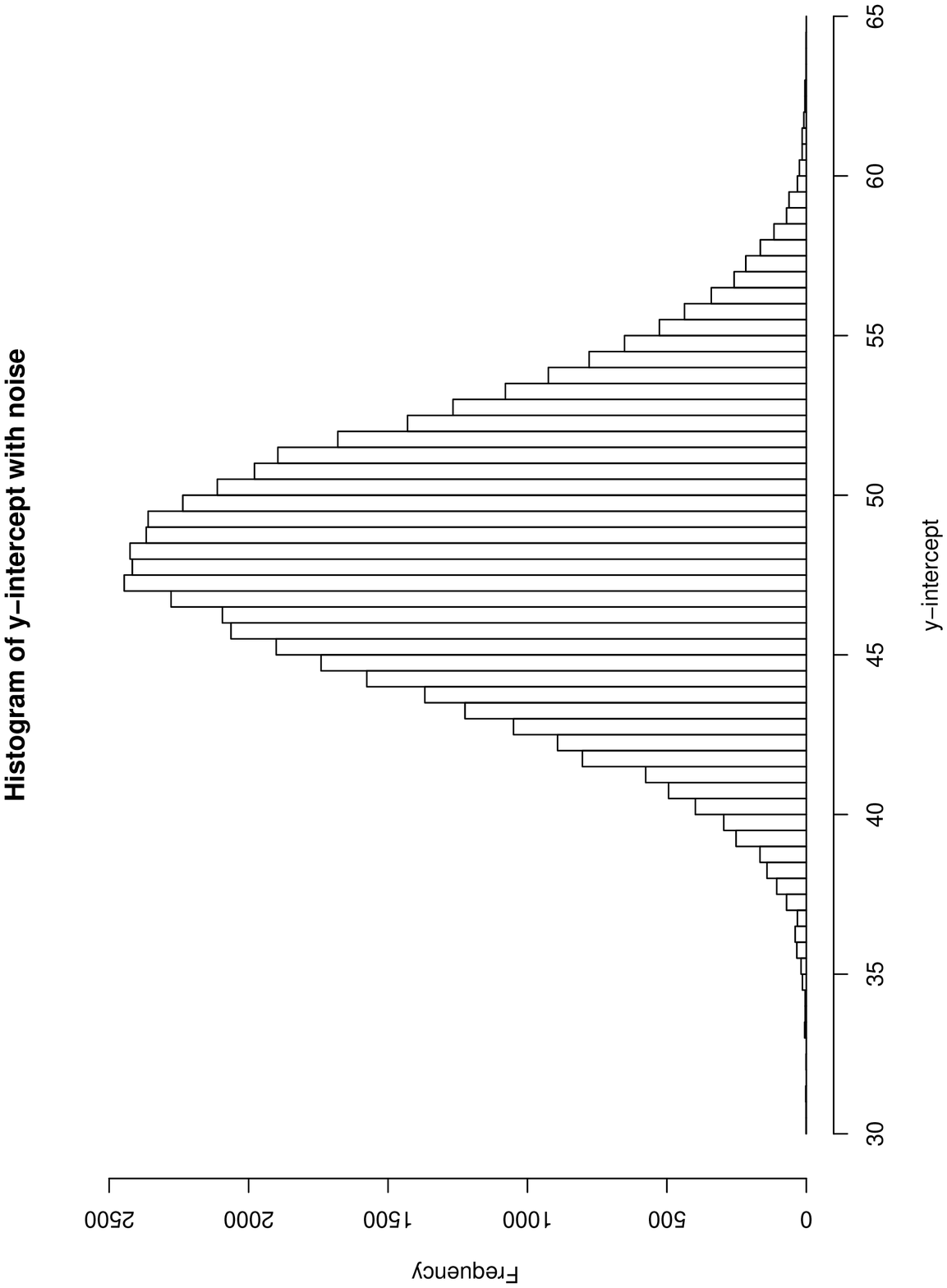, width=10cm}
  \end{turn}
  \end{center}
\caption{
Histogram of $G_0$ for human sex-averaged chromosome regression
$G_{ch, sex-ave, human} = G_0 +k P$ when noise is added to the genetic
length $G$. The noise was modeled as a normally distributed
variable with zero-mean and
standard deviation (sd) of 8.51517, which is the 
observed sd of regression residuals $\epsilon= G- G_0- kP$,
for $G_{ch,sex-ave,human} \sim P$ in Eq.(1).
}
\end{figure}

\newpage
%%%%%%%%%%%%%%%%%%%%%%%%%%%
\section{AIC and BIC of two-parameter regression models vs. one-parameter
models (Table A1) }

\begin{table}[ht]
\begin{center}
\begin{tabular}{r|c|c}
\hline
source of genetic length data  & $\Delta $AIC= AIC$_2-$AIC$_1$ & $\Delta $BIC= BIC$_2-$BIC$_1$  \\
\hline
human chromosome, sex-averaged & $-42.5$ & $-41.4$    \\
 female & $-45.3$ & $-44.2$  \\
 male & $-34.5$ & $-33.4$  \\
\hline
human ch. arm, sex-averaged  & $-52.9$ & $-51.4$  \\
 female  &   $-43.8$ & $-42.3$     \\
 male  &  $-43.6$ & $-42.1$  \\
\hline
rat, chromosome, sex-averaged & $-9.9$ & $-8.9$ \\
mouse, chromosome, sex-averaged & $-1.1$ & $-0.2$ \\
\hline
chicken, chromosome, sex-averaged & $-28.7$ & $-27.5$ \\
\hline
honeybee, chromosome, sex-averaged & 1.9 & 2.7 \\
 remove chromosome 1 & 1.0 & 1.7 \\
\hline
yeast, chromosome, sex-averaged & $-6.1$ & $-5.3$\\
\hline
worm,central region & 1.4 & 0.96 \\
 distal region & $-7.9$ & $-7.6$ \\
\hline
\end{tabular}
\end{center}
\caption{\label{tablea1}
Difference of Akaike and Bayesian information criterion (AIC and BIC)
between the two- and one-parameter regression models for modeling chromosome-scale
recombination in 7 genomes. A negative $\Delta AIC$ or $\Delta BIC$ value
favors the two-parameter regression model.
}
\end{table}

\newpage

\section{Chicken genetic length vs. physical length in log-log scale  (Fig.A4)}

\begin{figure}[ht]
 \begin{center}
  \begin{turn}{-90}
   \epsfig{file=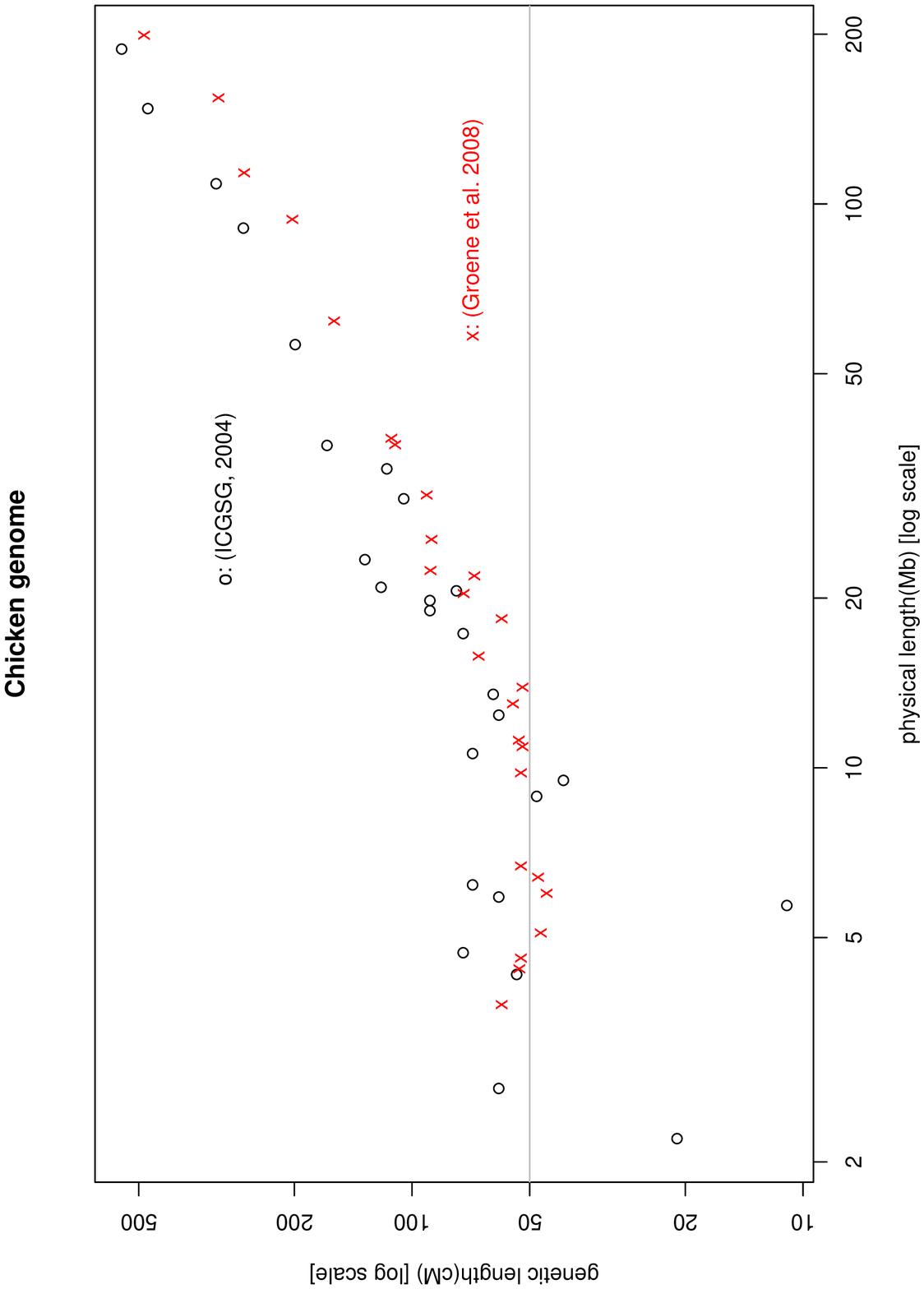, width=10cm}
  \end{turn}
 \end{center}
\caption{
The genetic length (in cM) vs. physical length (in Mb) plots for
chicken genome ({\sl Gallus gallus}) in log-log scale (for linear-linear
scale, see Fig.2(B)). 
% The genetic length of 50cM is marked  by a horizontal line.
}
\end{figure}

\newpage

\section{Two-parameter regression model of Opossum chromosome genetic length (Fig.A5)}

\begin{figure}[ht]
 \begin{center}
  \begin{turn}{-90}
   \epsfig{file=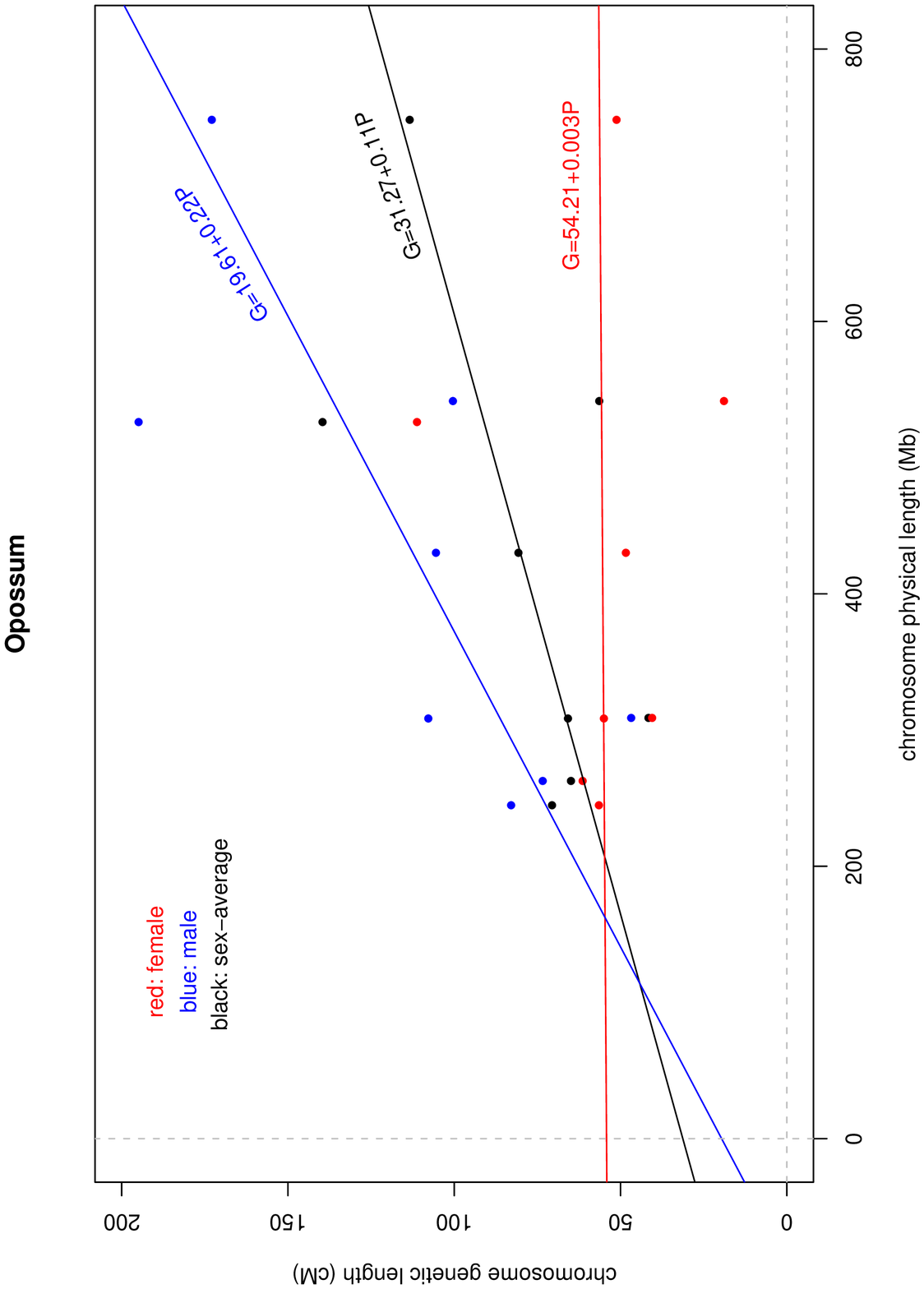, width=10cm}
  \end{turn}
 \end{center}
\caption{
Two-parameter regression of  Opossum {\sl Monodelphis domestica} 
chromosome genetic length (cM)
over physical length (Mb). The regression coefficients for female map (red)
are: $G_{female}= 54.206 + 0.003 P$, for male map (blue):
$G_{male}= 19.610 + 0.216 P$, and for sex-averaged map (black):
$G_{sex-ave}= 31.275 + 0.114 P$.
}
\end{figure}

\newpage

\section{Recombination rates of six genomes as a function of the smallest chromosome size (Fig.A6)}

\begin{figure}[ht]
 \begin{center}
  \begin{turn}{-90}
   \epsfig{file=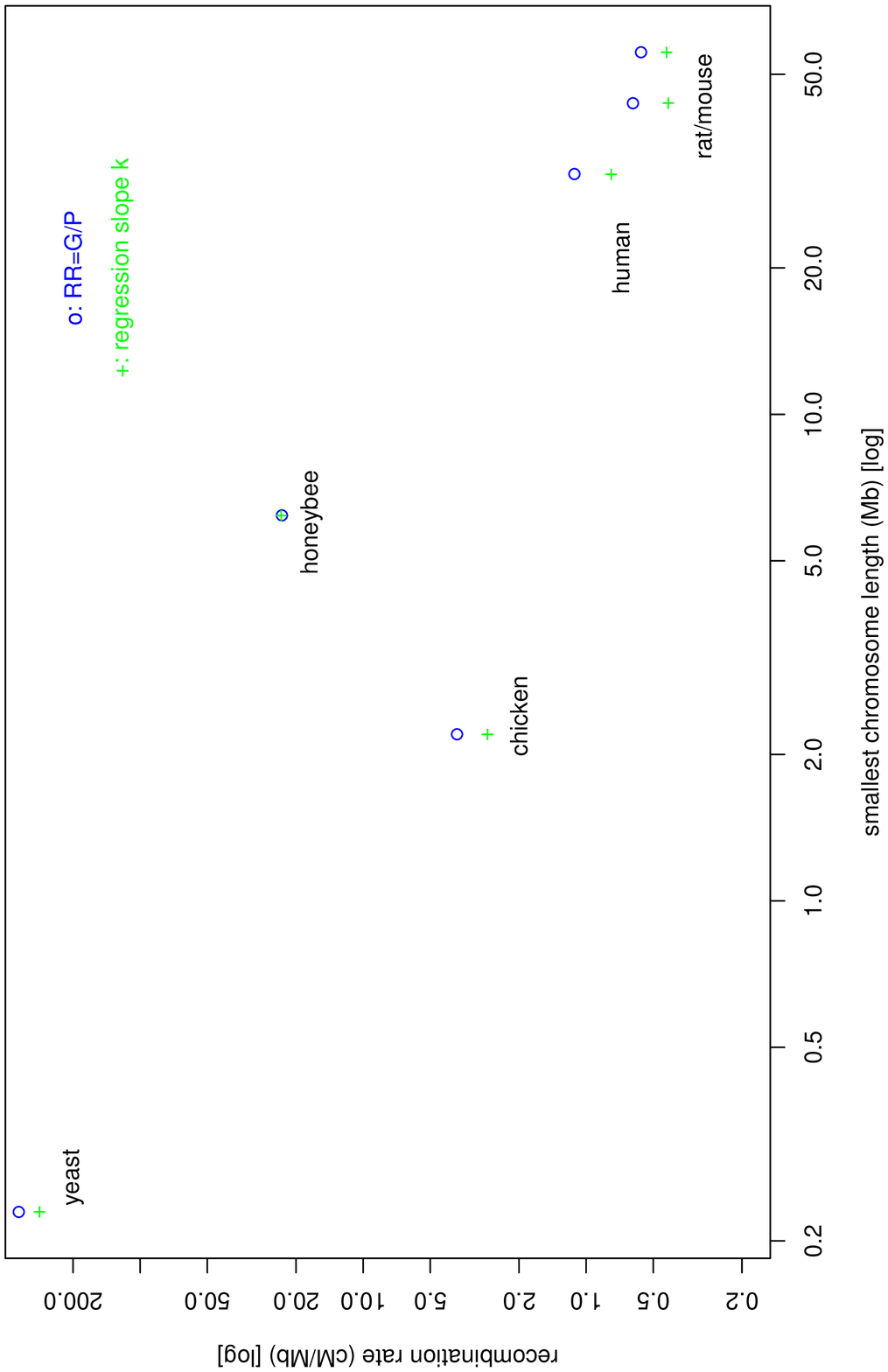, width=10cm}
  \end{turn}
 \end{center}
\caption{
Recombination rates of six genomes as a function of the smallest chromosome
size. Genome-wide recombination rate is measured both by the genome-wide genetic-to-physical
length ratio ($RR=G/P$) (circles) and by the regression slope $k$ (pluses).
The plot is in log-log scale.
}
\end{figure}

\newpage

\large

\end{document}